\documentclass[twocolumn,prl,showpacs,amsfonts,superscriptaddress,floatfix,nofootinbib]{revtex4-1}

\usepackage{amsmath}
\usepackage{graphicx}
\usepackage{footmisc}
\usepackage{color}

\newcommand{\nextnearsites}[1]{\langle\!\langle #1 \rangle\!\rangle}
\newcommand{\nearsites}[1]{\langle #1 \rangle}
\newcommand{\ee}{\mathrm{e}}
\newcommand{\ii}{\mathrm{i}}
\newcommand{\pdagger}{{\vphantom{\dagger}}}
\newcommand{\vect}[1]{{\mathbf{#1}}}

\newcommand{\eref}[1]{(\ref{#1})}

\let\veeOriginal\vee
\let\wedgeOriginal\wedge
\renewcommand{\vee}{\;\veeOriginal\;}
\renewcommand{\wedge}{\;\wedgeOriginal\;}

\begin{document}

\title {The Chern states on the honeycomb and Lieb lattices}

\author{Igor N. Karnaukhov and Igor O. Slieptsov}

\affiliation{G.V. Kurdyumov Institute for Metal Physics, 36 Vernadsky Boulevard, 03142 Kiev, Ukraine}

\begin{abstract}  {
The Haldane model of the Chern insulator is considered on the Lieb and honeycomb lattices. We provide a detailed analysis of the model's ground-state phase diagram and demonstrate a scenario of the topological phase transitions in the system with a single-particle spectrum that includes flat and dispersion bands, that is realized on the Lieb lattice. We find that the Chern number of the flat band is non-zero, depending on the parameters of the model. We define the topological metal state as an intermediate state between topological insulator and trivial metal. The phase transition between topological insulator and topological metal states is accompanied by continuous changing of the Chern number, a jump of the surface charge or spin current defines the point of the topological metal-trivial metal phase transition. The results have been illustrated with numerical calculations of the model.}
\end{abstract}
 \maketitle

\section{Introduction}

The Haldane model~\cite{Hal} is a key model for understanding the topological states of 2D fermion systems--- the Chern insulators. The Chern number $C$ is well defined in insulator state, being an integer. The~Berry curvature is a gauge-invariant for both insulator and metal phases, in metal phase it defines a Chern number as an integral over the Fermi surface (in contrast to insulator phase, where an integral is defined over the Brillouin zone). The topological state in the metal state may be characterized by the charge or spin Chern numbers, which define the surface charge or spin currents that have a topological nature in the metal state. We will study the Chern state in insulator and metal phases in the framework of the Haldane model defined on two different 2D lattices in order to investigate the peculiarities of their behavior.

Models of strongly correlated electrons in decorated 2D lattices (such as the Lieb lattice) are extensively studied, motivated by the search for metallic (flat-band) ferromagnetism~\cite{A}. Phase diagrams of such models include different phases with spin, charge and spin-charge orderings. The topological and nematic phases are realized on the Lieb lattice for single fermion states forming a flat band~\cite{1,1a}. Traditionally, the topological insulator's (TI) behavior derives from the dispersive bands that have nontrivial Chern numbers or $\mathbb{Z}_2$~indices. The systems with dispersionless (flat) bands can be topological nontrivial, e.~g. flat Landau bands for particles in a magnetic field, where the fractional quantum Hall effect is the result of nontrivial correlated physics \cite{g}.
A 2D electron gas in transverse magnetic field, known as the Hofstadter model~\cite{Hof}, represents various quantum Hall states, each one is characterized by a quantum number--- the Chern number~\cite{Haf}.  The Chern insulator state where the bulk-edge correspondence accompanies the topological insulator state is realized in the Hofstadter model~\cite{Hof,c} (see also ref.~\cite{d}).

A model of TI on the Lieb lattice in 2D and its 3D counterpart the perovskite or edge centered cubic lattice that takes in account a spin-orbit interaction term has been presented in ref.~\cite{1}. The next-nearest neighbor spin-orbit-induced interaction opens a gap in the $M$ point and the system is TI at 1/3 and 2/3 band filling with $\mathbb{Z}_2$ index equals to 1 for two dispersive bands and 0 for topologically trivial flat band. In ref.~\cite{1a} the authors have considered a variant of the tight-binding model on the Lieb lattice that takes into account a staggered potential, nearest-neighbor (NN) and next-nearest-neighbor (NNN) hoppings of fermions. They have shown that the point of band crossing is not topologically nontrivial, but also only weak interaction can induce the TI phase. As~a~rule, the Chern number of the flat bands is $C_{\mathrm{flat}}=0$, therefore (opposed to Landau levels) we get no quantized Hall conductance for these particular flat bands.

We consider topological phase transitions in the framework of the Haldane model~\cite{Hal} on the honeycomb and Lieb lattices. The studied model is a variant of the Haldane model realizing on the Lieb lattice --- the tight-binding model of TI with the NN and NNN hoppings. The orbitals on the different sites of unit cell in the cubic lattice have different energies.  The phase diagram of the model also depends on the filling of the fermion bands, the TI phases are realized at 1/3 and 2/3 filling when the Fermi level is positioned within a forbidden band between flat and neighboring dispersing bands. At the point of the topological phase transition (TPT) the flat band and a neighboring dispersing band touch at the Dirac point. Breaking the time-reversal symmetry (TRS) leads to topological states that are fundamentally different from traditional insulator phases. In the studied model an inversion symmetry is conserved, the system exhibits a topological Chern insulator.

The purpose of this paper is to understand the topological state of the fermion system with different spectrum (namely band and flat band spectrum), to calculate the Chern numbers in both insulator and metal phases. We~will shown also that the flat band is isolated in the TI phase due to breaking TRS, therefore the flat band can be viewed as critical point at TPT.
In other words, we correlate the topological nature of the flat band and TPT from which emerges a reincarnation of  the quantum Hall conductivity. We~will show that the Chern number of the flat band is changed from zero to one depending on parameters of the model. The Chern state exhibits the chiral gapless edge modes that define the surface charge or spin currents in metal and insulator phases.

\section{The studied 2D model of the Chern insulator}
\begin{figure}[tbp]
    \centering{\leavevmode}
    \begin{minipage}[h]{0.4\linewidth}\center{\includegraphics[width=\linewidth]{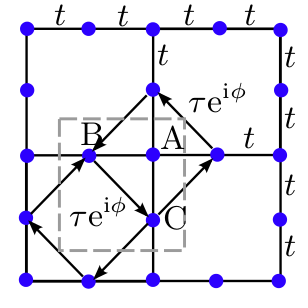} \\ (a)}\end{minipage}
    \begin{minipage}[h]{0.45\linewidth}\center{\includegraphics[width=\linewidth]{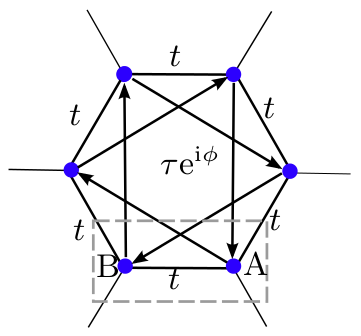} \\ (b)}\end{minipage}
    \caption{
        (a)~Schematic plot of the Haldane model on the Lieb lattice with the real nearest $t$ and complex (with chosen direction) next-nearest hoppings $\tau \exp(i\phi)$, a dotted curve indicates the unit cell, which contains three sublattice sites $(A,B,C)$,
        (b)~the~Haldane model on the honeycomb lattice.
    } \label{fig:Model}
\end{figure}

We will analyze the tight-binding model of the TI defined on the honeycomb and Lieb lattices, that describes an influence of a magnetic field. In the real space the Hamiltonian of the model is
\begin{equation}
    {\cal H} = -t\sum_{\nearsites{i,j}} a^\dagger_{i} a^\pdagger_{j} -\tau \sum_{\nextnearsites{i,j}} \ee^{\ii \phi_{ij}} a^\dagger_{i}a^\pdagger_{j} +
        \epsilon \sum_{j=A} a^\dagger_{j} a^\pdagger_{j},
    \label{eq:H0}
\end{equation}
where $a^\pdagger_{j} $ and $a_{j}^\dagger$ are the spinless fermion operators with the usual anticommutation relations.
The first term represents NN hopping with a magnitude~$t$, the second term is NNN hopping with a hopping parameter $\tau$ and a Peierls phase~$\phi_{ij}$, the last term represents a staggered potential with a value~$\epsilon$. The unit cell of the Lieb lattice consists of three sites for the lattice denoted as $A$, $B$, $C$ and two sites $A$ and $B$ for the honeycomb lattice (see fig.~\ref{fig:Model}). The homogeneous Peierls phase $\phi_{ij}$ (denoted as $\pm\phi$ the clockwise (anticlockwise) NNN hopping relatively to a cell) is considered as the parameter of the model.
The model~(\ref{eq:H0}) defined on the honeycomb lattice has been introduced by Haldane~\cite{Hal}.
The phase state of the model on the  Lieb lattice for two different complex phases $\phi=\pi/4$ and $\phi=\pi/2$ at filling 1/3 has been investigated recently in ref.~\cite{2D}.

The NNN hoppings in~(\ref{eq:H0}) breaks TRS which is crucial in obtaining the TI phase. The staggered potential does not break inversion symmetry, the spectrum of one-particle excitations is a spatial inversion symmetric.
We~will investigate the ground-state phase diagram in a low filling region at ${0<n\leq 1/2}$. Using the particle-hole transformation $a_{j}^{\dagger} (a_{j}^\pdagger)\to a_{j}^\pdagger(a_{j}^{\dagger})$ for the Hamiltonian (\ref{eq:H0}), one may obtain the phase diagram at filling $1/2<n\leq 1$, as ${\cal H}(t,\tau,\phi,\epsilon,n)\to {\cal H}(-t,-\tau,\phi,-\epsilon,1-n)$. The model (\ref{eq:H0}) can be considered as the Haldane model defined on the Lieb and honeycomb lattices.

\section{The honeycomb lattice}

\begin{figure}[tb]
    \centering{\leavevmode}
    \begin{minipage}[h]{0.48\linewidth}\center{\includegraphics[width=\linewidth]{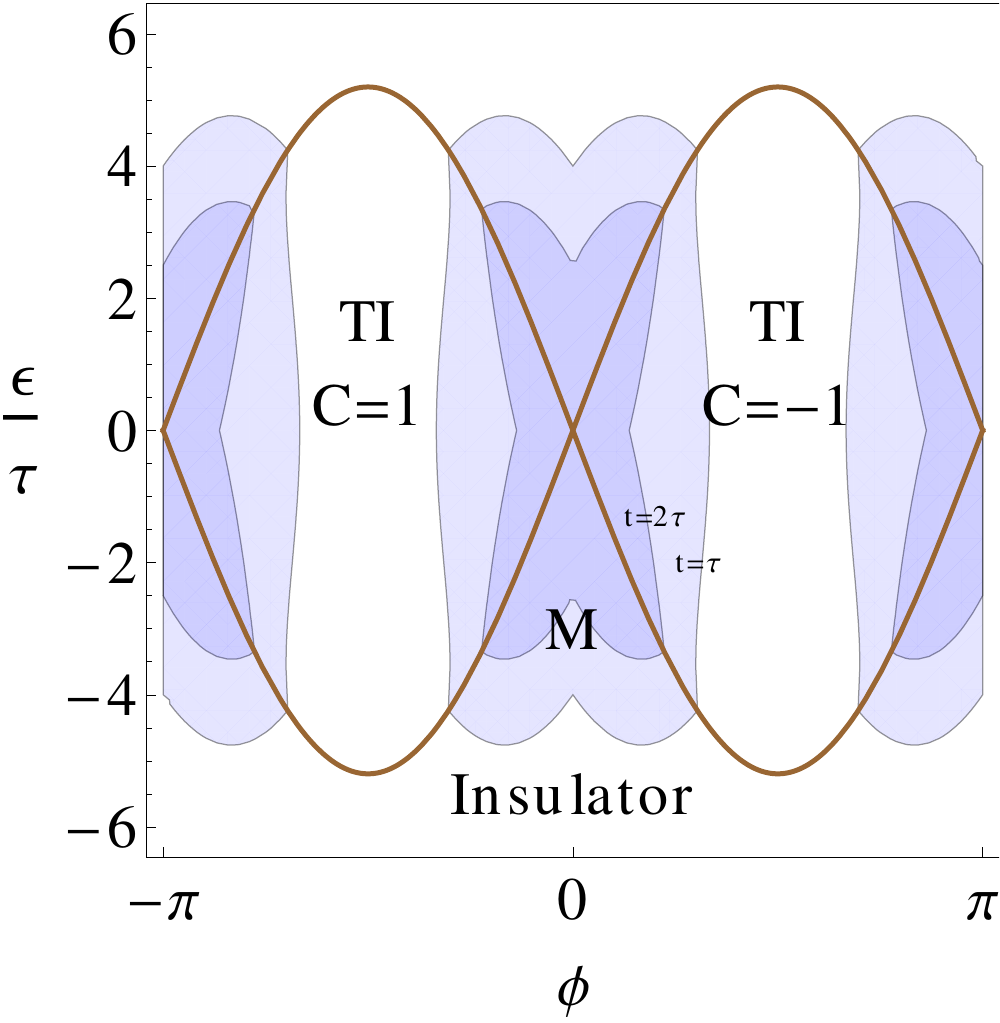} \\ (a)}\end{minipage}
    \begin{minipage}[h]{0.48\linewidth}\center{\includegraphics[width=\linewidth]{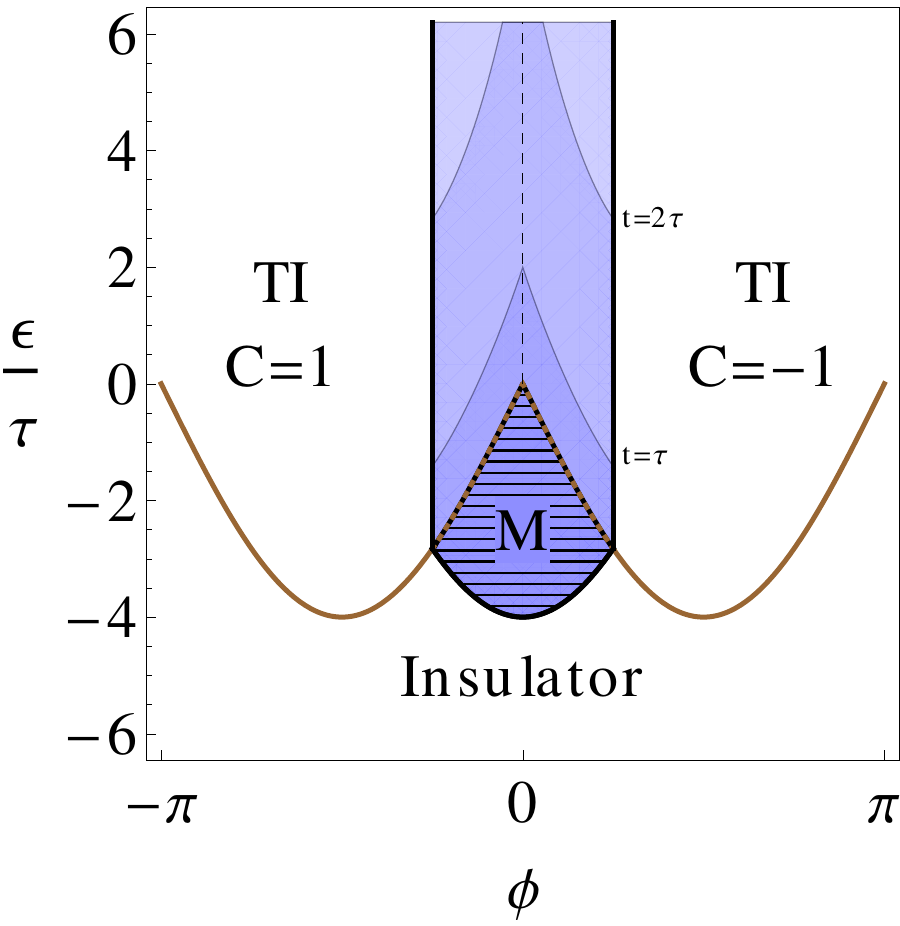} \\ (b)}\end{minipage}
    \caption{
        Phase diagrams of (a) the Haldane model on the honeycomb lattice at half-filling ${n=1/2}$ and (b) the model on the Lieb lattice~\eref{eq:H0} at 1/3 band filling with three distinct phases: metal~(M), insulator and topological insulator~(TI).
        The topological phase transition is marked with a brown line (a) ${\epsilon = \pm3\sqrt3 \tau\sin\phi}$ and (b) ${\epsilon = -4\tau\sin\phi}$.
        Blue regions stand for metal phase. The shaded region represents metal phase for arbitrary (including infinitesimal) values of~$t$.
        The black curve limits metal phase for $t\gg\tau,\epsilon$.
        Two lines separating metal (inside) and TI (outside) phases are shown for $t=\tau$ and $t=2\tau$.
    } \label{fig:HPD} \label{fig:MI}
\end{figure}

The phase diagram of the Haldane model has a rich structure for different values of $t$ and $\tau$~\cite{I2}, in the case $t\gg\tau$ the system is in the insulator state at half-filling ${n=1}$. For arbitrary values of the parameters of the Hamiltonian (\ref{eq:H0}) the phase diagram consists of insulator, TI and metal phases at ${n=1}$. A traditional Haldane diagram $\epsilon =\pm 3\sqrt3 \tau \sin \phi$ (brown lines in fig.~\ref{fig:HPD}) that separates insulator and TI phases is realized in the $\tau\ll t$ limit. At $t<5\tau$ the phase diagram contains a metal phase which is shown in fig.~\ref{fig:HPD} for different values of $t$, this region decreases with increasing~$t$.

\begin{figure}[tb]
    \centering{\leavevmode}
    \begin{minipage}[h]{0.48\linewidth}\center{\includegraphics[width=\linewidth]{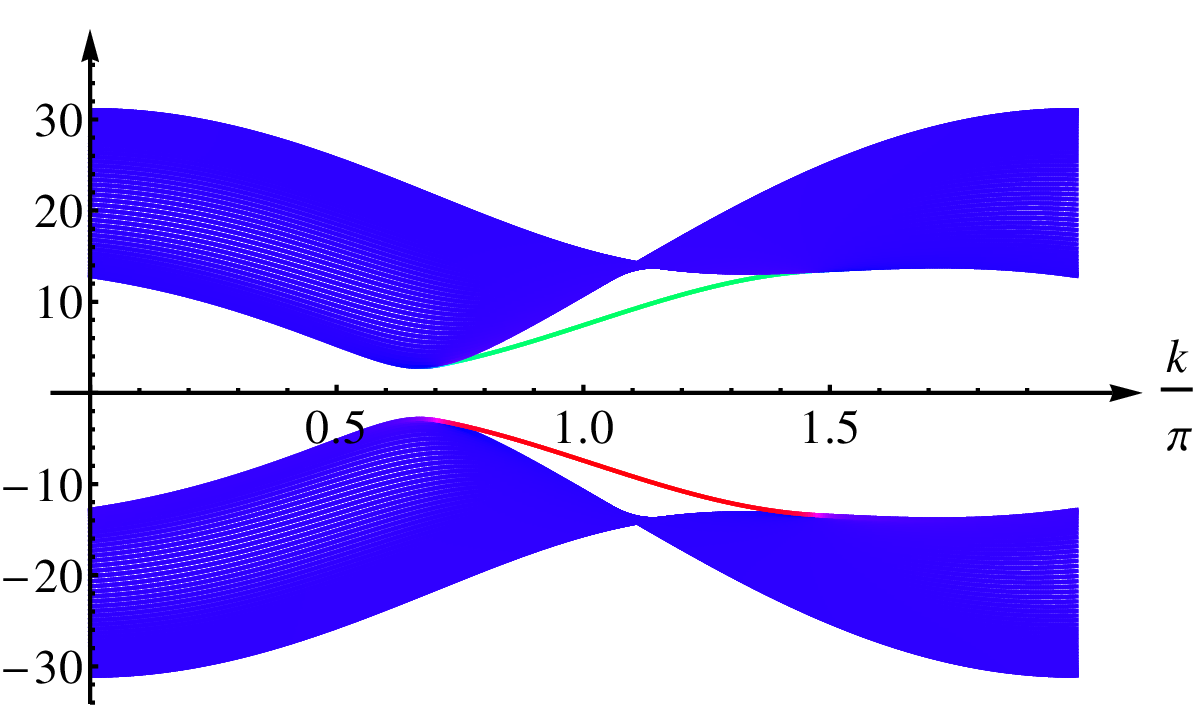} \\ (a)}\end{minipage}
    \begin{minipage}[h]{0.48\linewidth}\center{\includegraphics[width=\linewidth]{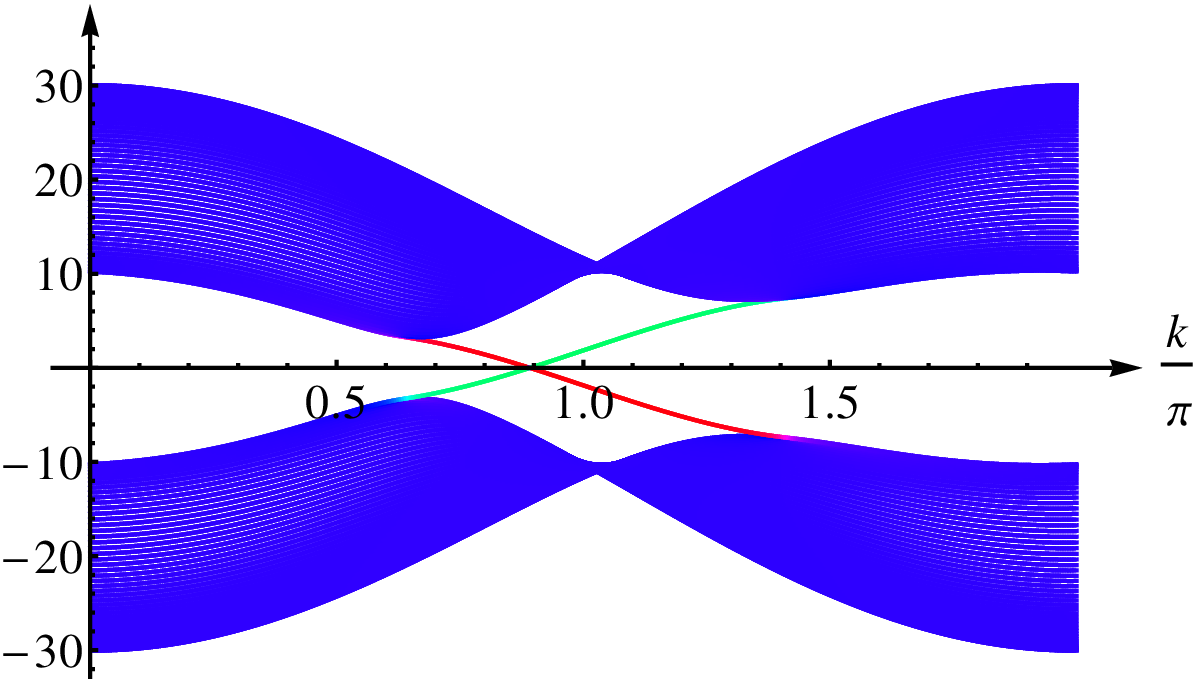} \\ (b)}\end{minipage}
    \caption{
        Energy levels of the Haldane model on the honeycomb lattice calculated on a cylinder with open boundary conditions for a zig-zag boundary (a) in the (topologically trivial) insulator state at ${t=10\tau}$, ${\phi=\pi/2}$, ${\epsilon=6\tau}$ and (b) in TI state at ${t=10\tau}$, ${\phi=\pi/2}$, ${\epsilon=2\tau}$. The wave vector $k$ is along the boundary.
    } \label{fig:HSpec}
\end{figure}

The phase of TI is defined by the Haldane diagram at ${t\gg\tau}$, the spectrum of excitations has a traditional form for TI (see in fig.~\ref{fig:HSpec}), where two bulk fermion subbands are connected via edge chiral modes. The Chern number and chiral edge modes define the state of the Chern insulator.
In the system with zero-net magnetic field~\cite{Hal} a filled band ($\gamma$ is a band index) with nontrivial Chern number $C_\gamma$ yields a Hall surface conductance ${\sigma_H = C_\gamma e^2/h}$. The Chern number is a topological invariant which can be easily defined for a band isolated from all other bands by the formula
\begin{equation}
    C_\gamma =\frac{1}{2\pi}\int_{BZ}{\cal B}_\gamma (\vect{k}) \mathrm{d}^2\vect{k}
    \label{eq:Chern}
\end{equation}
integrating over the Brillouin zone (BZ) of the system, where ${\cal B}_\gamma (\vect{k}) = \nabla_{\vect{k}}\times {\cal A}_\gamma (\vect{k})$ is the Berry curvature.
The~Berry potential
$${\cal A}_\gamma (\textbf{k})= -\ii \langle u(\vect{k}) | \nabla_\vect{k} | u(\vect{k}) \rangle = -\ii \sum_\rho^N \left(u^{ \gamma}_\rho(\vect{k})\right)^* \nabla_{\vect{k}}u^{ \gamma}_\rho (\vect{k})$$
is defined in terms of the Bloch states $u_\rho^\gamma (\textbf{k})$, $\rho$ is the index of the unit cell, $N$ is the total number of unit cells. The change of the value of the Chern number is realized at the point of the TPT in which the linear bands are reminiscent of a Dirac-like point.

\begin{figure}[tb]
    \centering{\leavevmode}
    \begin{minipage}[h]{0.45\linewidth}\center{\includegraphics[width=\linewidth]{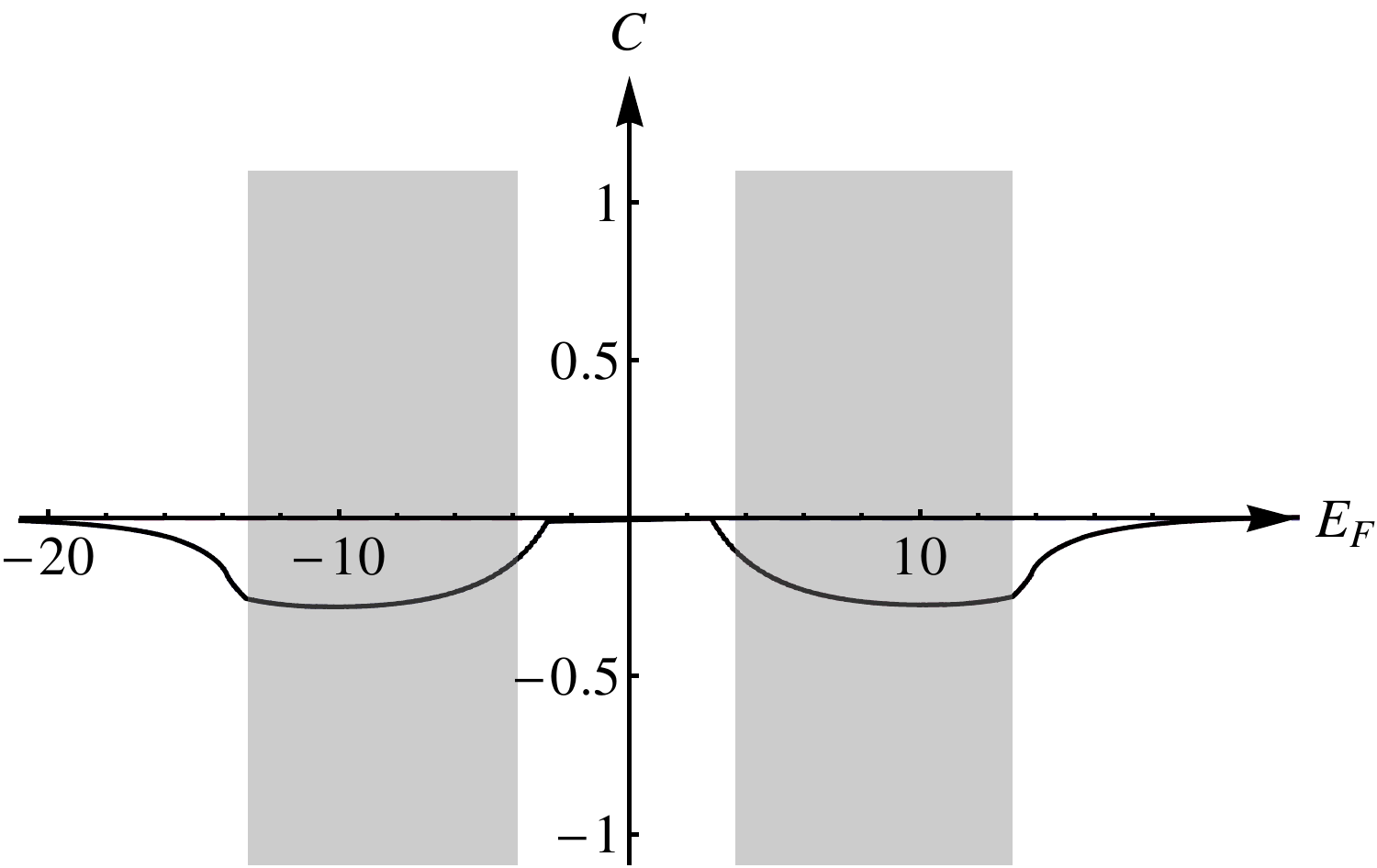} \\ (a)}\end{minipage}
    \begin{minipage}[h]{0.45\linewidth}\center{\includegraphics[width=\linewidth]{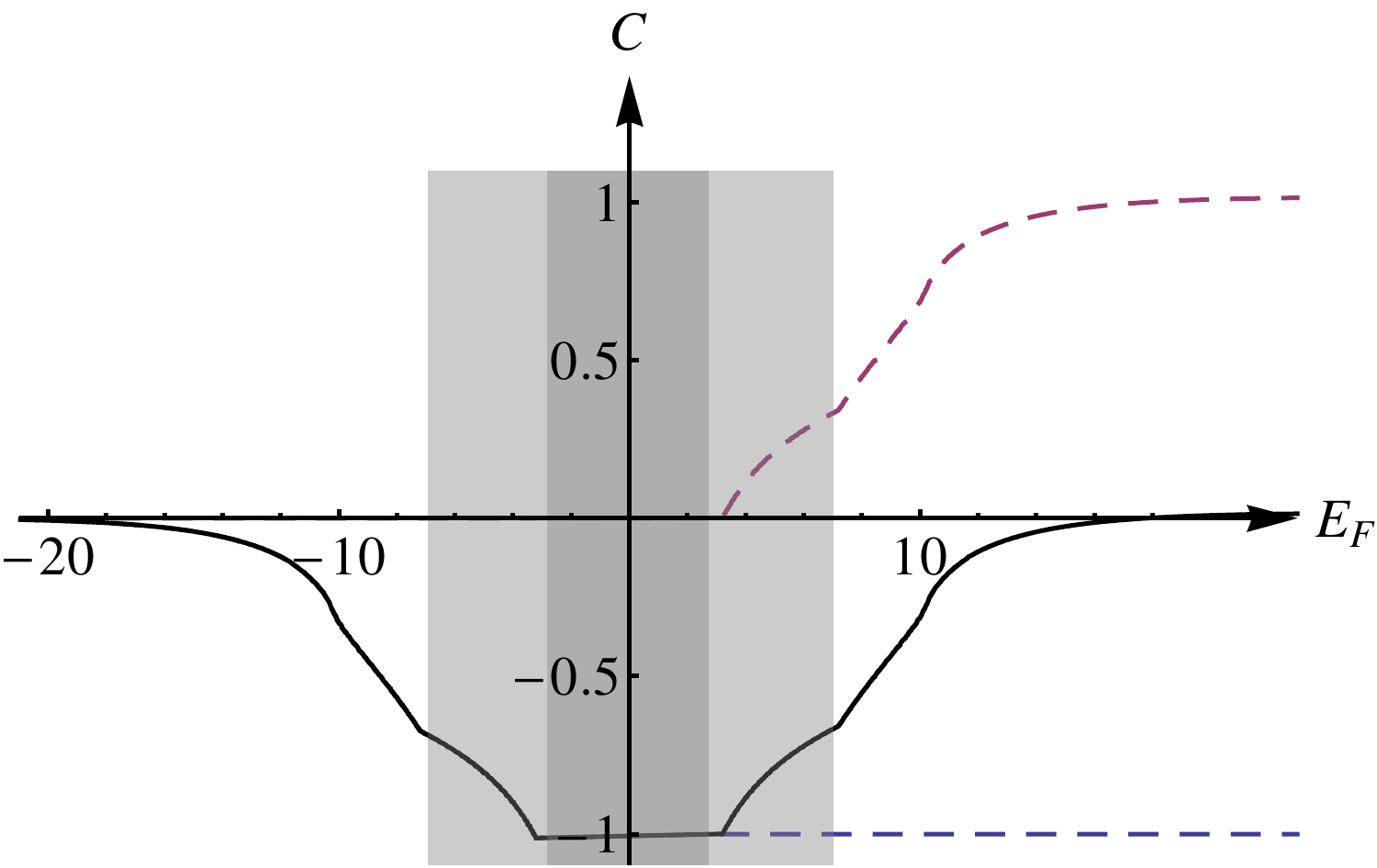} \\ (b)}\end{minipage}
    \caption{
        The Chern number of the Haldane model as function of the Fermi energy $E_F$ calculated  at (a) ${t=10 \tau}$, ${\phi=\pi/2}$, ${\epsilon=6\tau}$ and (b) ${t=10\tau}$, ${\phi=\pi/2}$, ${\epsilon=2\tau}$ (the same parameters as in fig.~\ref{fig:HSpec}), upper (red) dashed line stands for an upper subband, lower (blue) dashed line is for lower subband, a solid black line is for the total Chern number.
    } \label{fig:CH}
\end{figure}

The Chern number (\ref{eq:Chern}) is well defined in the insulator state, due to the bulk-edge correspondence two edge chiral modes define the surface Hall conductance. The~surface Hall conductance is also realized in the metal state for different filling, when the Fermi level crosses the edge chiral modes.
The~surface conductance is defined by the number of the chiral edges, that crosses at one point the Fermi surface.
Let~us consider the behavior of the system in the topological state for given parameters of the model (${t=10 \tau}$, ${\phi=\pi/2}$, ${\epsilon=2\tau}$) for different filling, namely the dependence of the Chern number as function of the Fermi energy $E_F$. We have analyzed the dependence of the Chern number for different fillings, integrating Berry curvature over the Fermi surface at given filling. It~is convenient to calculate the Chern number for different values of $E_F$ (see figs~\ref{fig:CH}). The~transition from TI state (shady region at small values of~$E_F$) with the Chern number equal to $-1$ to the topologically trivial metal state (white region) with ${C=0}$ at large values of $E_F$ is realized via an intermediate state named as topological metal state (grey region) with ${0<C<1}$ and one gapless edge chiral mode (see fig~\ref{fig:CH}b). Two shown regions in figs~\ref{fig:CH} (namely shady and grey) are defined by two and one one gapless edge chiral modes, correspondently. The~topological metal phase is observed in the case of trivial insulator for $n\neq 1$ for a wide region of the Fermi energies (a grey region in fig~\ref{fig:CH}a).
In~other words, the topological metal state is characterized by a surface chiral edge current and ${0<C<1}$.

\section{The Lieb lattice or a line-centered square lattice}

In a traditional tight-binding  model on the Lieb lattice (the Hamiltonian (\ref{eq:H0}) with the nearest-neighbor hopping only) a flat band touches two linearly dispersing bands at the same point, where the linear bands are reminiscent of a Dirac-like point~\cite{A,1}. There are the three bands including a flat band with the dispersion at $\tau =0$
$$E_{\rm flat}(\textbf{k}) =\epsilon,$$ $$ \qquad E_\pm (\textbf{k}) = \epsilon \pm 2 t \sqrt{\cos^2(k_x/2)+\cos^2(k_y/2)},$$ at the $M$ point $E_\pm (\textbf{k}) \simeq |k|$, where $\textbf{k}=(k_x,k_y)$ is the wave vector.
The other two bands ($E_\pm$) describe the cone at the $M$ point when $\tau=0$,
and two parabolic bands separated from the flat band by the gaps
${\Delta_\pm=\epsilon \pm 4 \tau  \sin\phi}$, when $\tau, \epsilon \neq 0$.

\begin{figure*}[tb]
   \centering{\leavevmode}
    \begin{minipage}[h]{0.3\linewidth}\center{\includegraphics[width=\linewidth]{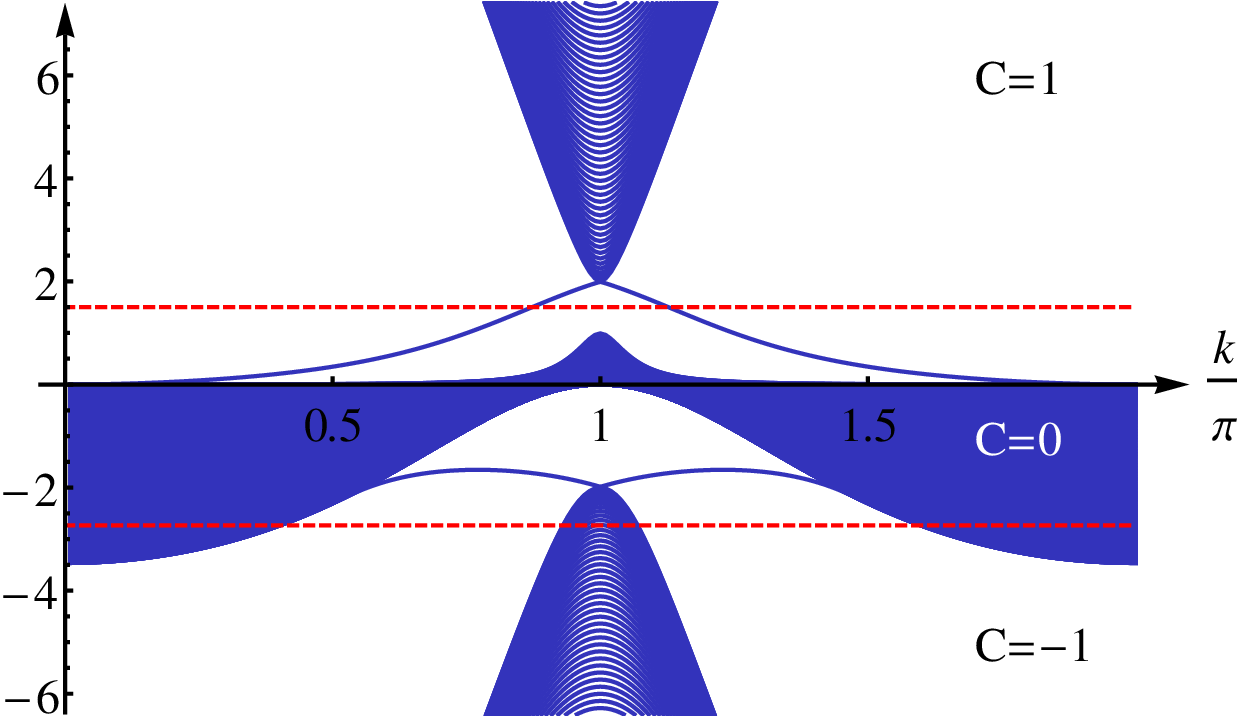} \\ (a)}\end{minipage}
    \begin{minipage}[h]{0.3\linewidth}\center{\includegraphics[width=\linewidth]{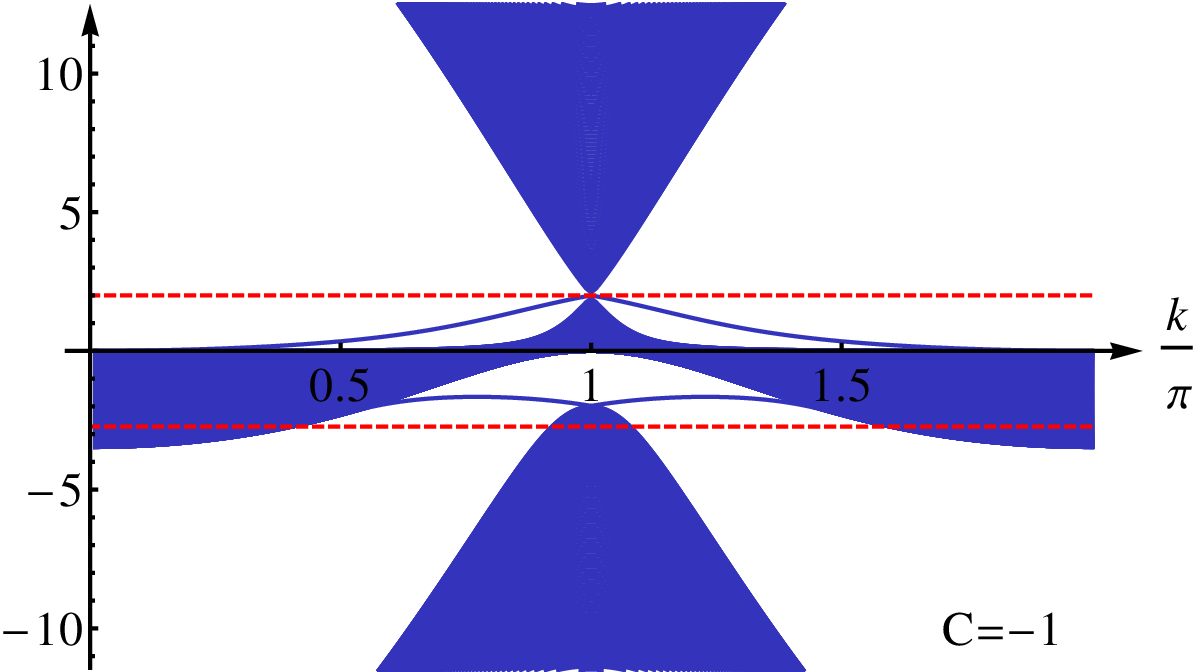} \\ (b)}\end{minipage}
    \begin{minipage}[h]{0.3\linewidth}\center{\includegraphics[width=\linewidth]{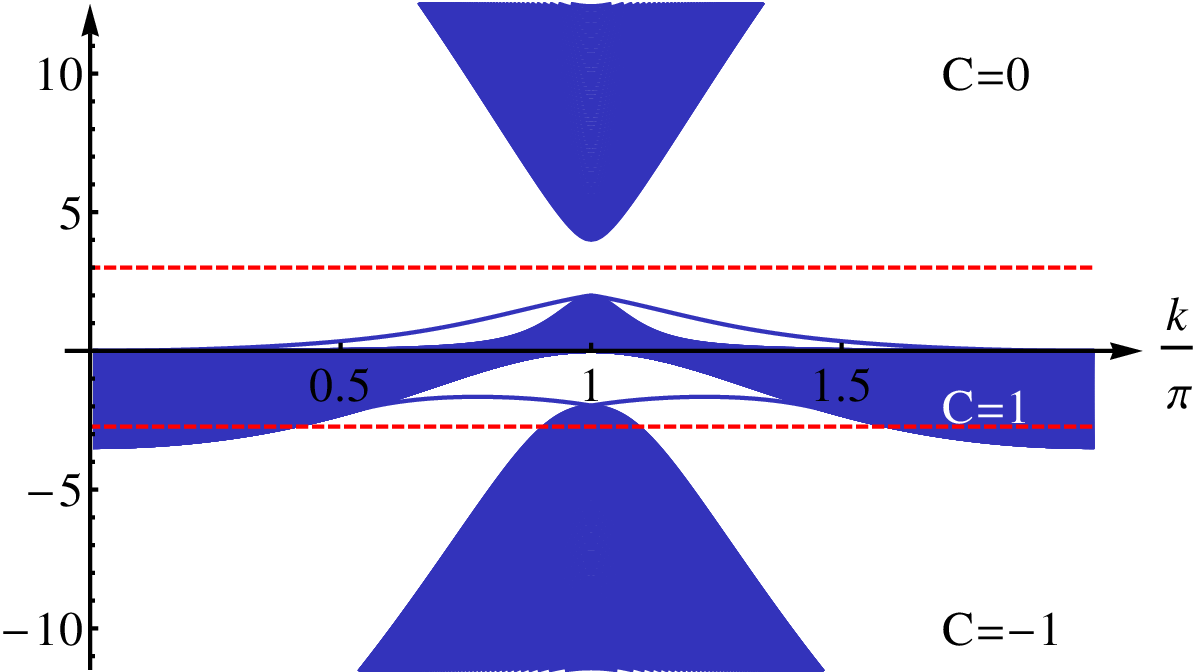} \\ (c)}\end{minipage}
    \begin{minipage}[h]{0.45\linewidth}\center{\includegraphics[width=\linewidth]{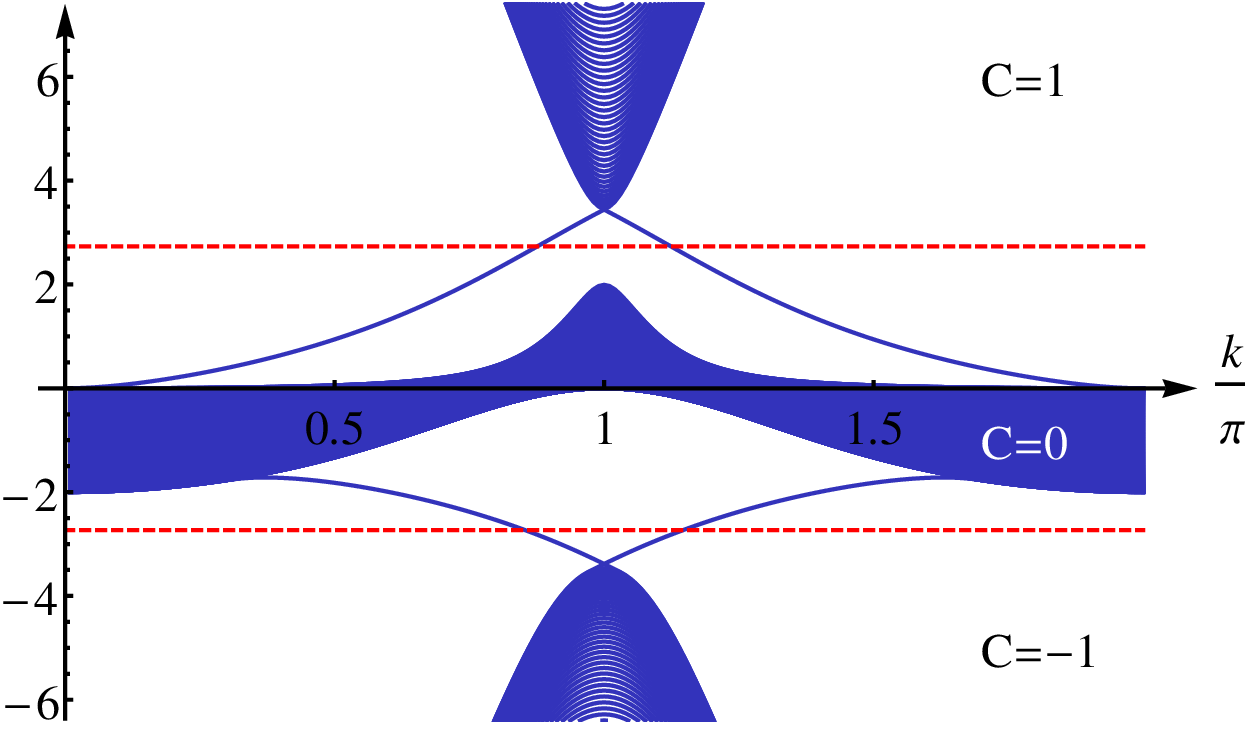} \\ (d)}\end{minipage}
    \begin{minipage}[h]{0.45\linewidth}\center{\includegraphics[width=\linewidth]{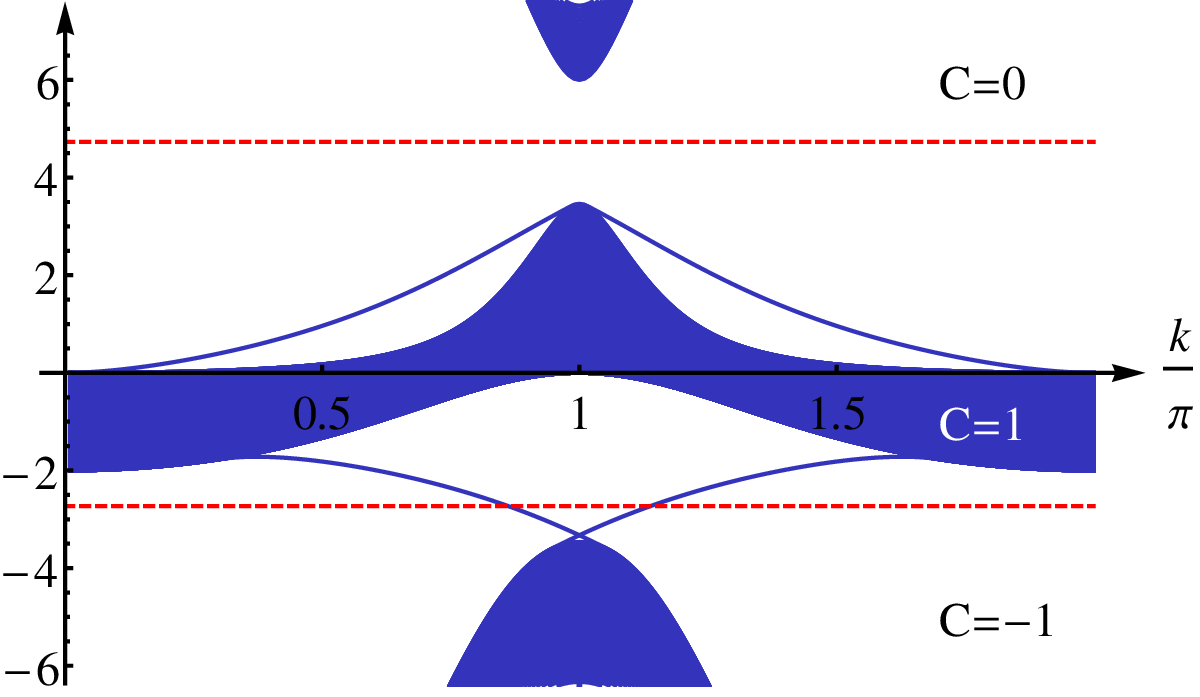} \\ (e)}\end{minipage}
    \caption{
      Dispersion curves $E(k)$ ($k$ is the wave vector along the boundary) and Chern numbers~$C$ of each band of the Lieb lattice finite in $y$-direction for different values of parameters, ${t=10 \tau}$,
      (a)~${\epsilon=  \tau}$, ${\phi=\pi/6}$,
      (b)~${\epsilon=2 \tau}$, ${\phi=\pi/6}$,
      (c)~${\epsilon=4 \tau}$, ${\phi=\pi/6}$,
      (d)~${\epsilon=2 \tau}$, ${\phi=\pi/3}$,
      (e)~${\epsilon=6 \tau}$, ${\phi=\pi/3}$.
     Fermi levels for ${n=1/3}$ and ${n=2/3}$ are marked with red dashed lines.
    } \label{fig:dispersion2D}
\end{figure*}

The flat band can be viewed as a critical point. In the case when $\tau$ and $\epsilon$ are  non-zero the flat band is isolated from the dispersing bands, it becomes slightly dispersive (without removing the band crossing point).

We consider the phase state depending on the model's parameters and focus our attention on the 1/3 band filling. The case $n=2/3$ is similar due to particle-hole symmetry.

\begin{figure}[tb]
    \centering{\leavevmode}
    \begin{minipage}[h]{0.8\linewidth}\center{\includegraphics[width=\linewidth]{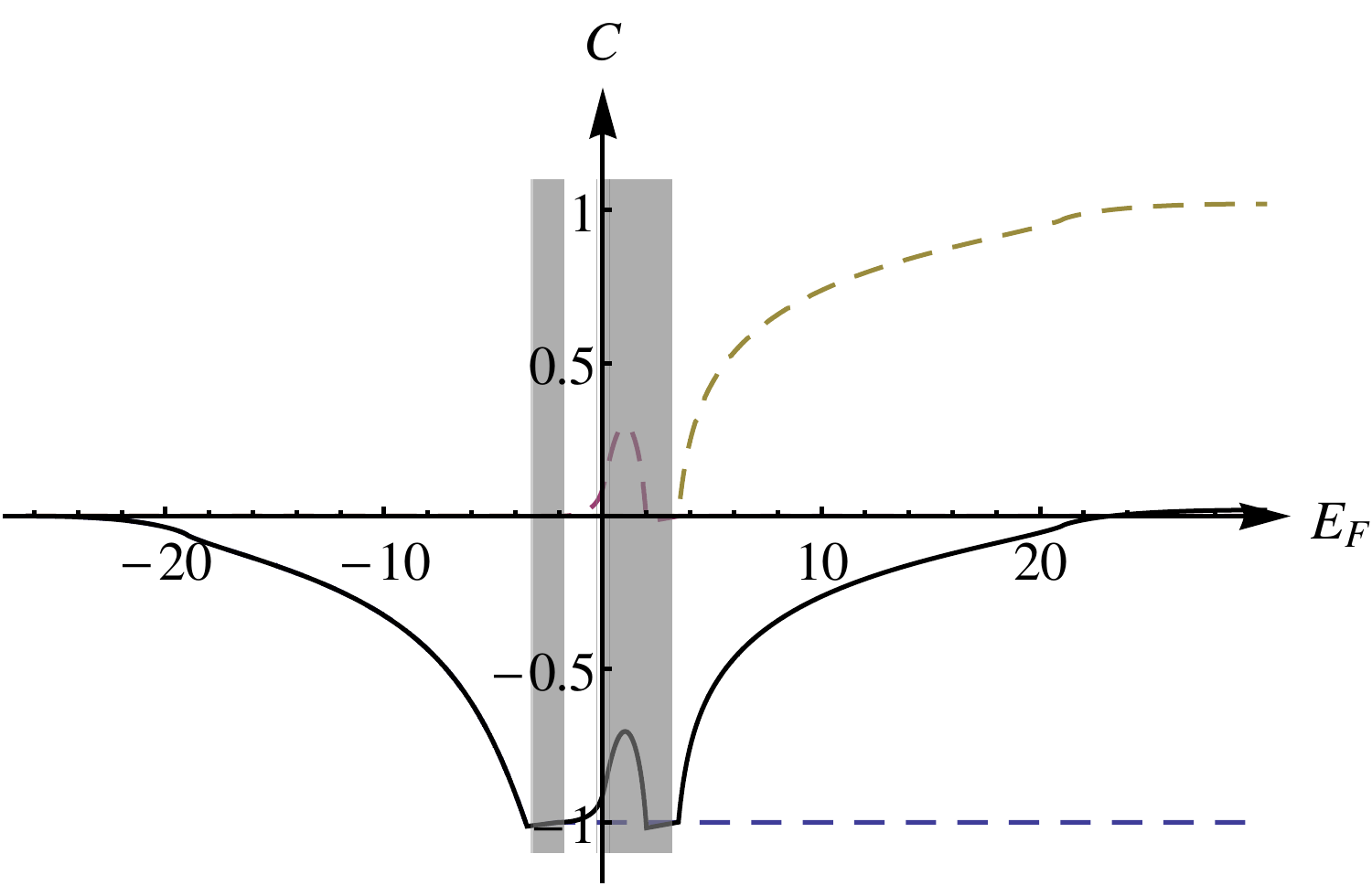} \\ (a)}\end{minipage}
    \begin{minipage}[h]{0.8\linewidth}\center{\includegraphics[width=\linewidth]{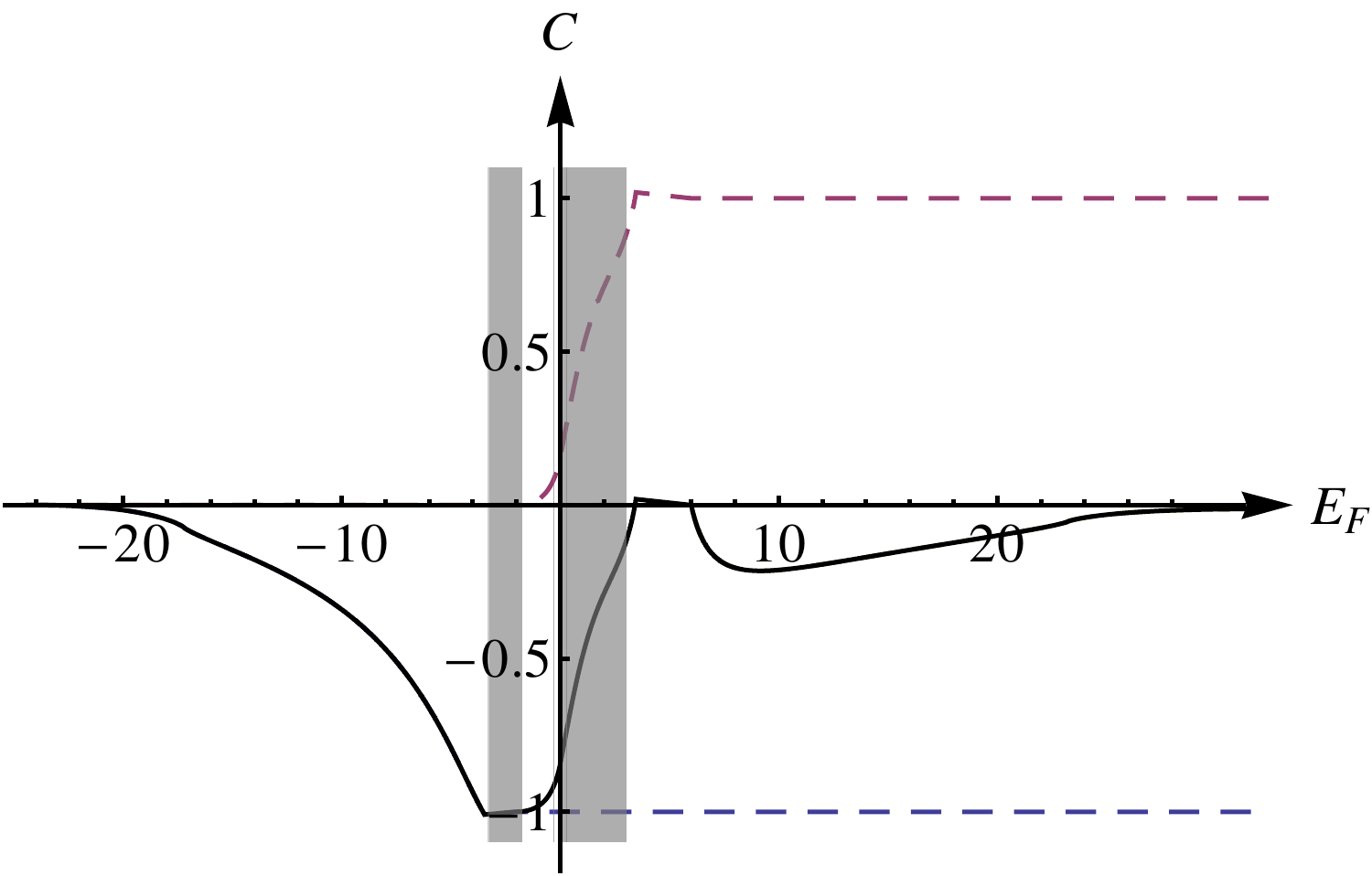} \\ (b)}\end{minipage}
    \caption{
       The Chern number as function of the Fermi energy $E_F$ numerically calculated on the Lieb lattice for (a) topologically trivial (${\epsilon =2\tau}$) and (b) topologically nontrivial (${\epsilon =6\tau}$) flat band, ${t=10\tau}$, ${\phi=\pi/3}$.
       Dashed lines stand for Chern number of each subband, while a solid line is a total.
    } \label{fig:CL}
\end{figure}

The phase transition to insulator phase can be realized at filling ${n=1/3}$ (${n=2/3}$). In~fig.~\ref{fig:MI}b the phase diagram describes the phase transition between metal (or bad metal) and insulator phase states at ${n=1/3}$. The Haldane-like diagram $\Delta_\pm =0 \iff \epsilon= \pm 4 \tau \sin \phi$ separates the phases of topological and trivial insulators for $\pi/4<\phi<\tfrac74\pi$, the Chern number calculated according to~\eref{eq:Chern} in the TI phase is equal to the signum of~$\phi$. The~metal phase is realized only when $-\pi/4<\phi<\pi/4$, a curve $\epsilon=-4 \tau \cos \phi$ separates this phase from the trivial insulator state. The region of metal phase depends on the ration of $t/\tau$ and reduces to a stripe in the infinite $t$-limit; it shrinks with $t/\tau$ decreasing. Most TIs are narrow band gap semiconductors, thus ${\tau/t\ll 1}$ in practical situations. The Haldane-like diagram is defined by the TPT with the Dirac-like one-particle spectrum at the point~$M$. Figs~\ref{fig:dispersion2D} demonstrate the one-particle spectrum of a finite 2D lattice and calculations of the Chern number of the subbands for different points of the phase diagram~Fig.\ref{fig:MI}.
The~one-particle spectrum has gaps in topologically trivial and nontrivial insulator phases and a ``quasiflat'' band touches dispersing band at a Dirac-like point.
The~insulating phase state is defined by the Chern number of each subband of spinless fermions below the Fermi level.
The~sign of the Chern number determines the chirality of the edge modes on the Fermi level.
As~we have mentioned, the middle and upper bands acquire Chern numbers plus and minus one in trivial insulator phase at $n=1/3$, while the lower one has $C=0$. Moreover, the flatness and the Chern number of the middle (``quasiflat'') band $C_{\rm flat}$ depend on the values of $\tau$, $\epsilon$ and $\phi$ and defined as
$$
  C_{\mathrm{flat}} = \left\{
    \begin{array}{ll}
        0, & |\epsilon|< 4 |\tau \sin\phi|, \\
        \operatorname{sgn} (\tau \sin\phi), & |\epsilon|> 4 |\tau \sin\phi|.
    \end{array}
  \right.
$$
In other words, the Haldane-like diagram is also strongly connected with the changing of the flat band's topology.

The edge states are symmetric, when fermions on the boundary interact with fermions along the boundary. We compare the calculations of the Chern number and the number edge modes that cross the Fermi surface for given filling of the system.  Such approach gives possibility to find ``bulk-edge correspondence in metallic phase'' in the model with a flat band. The behavior of the Chern number calculated on the Lieb lattice (figs~\ref{fig:CL}) for different $E_F$ is similar to the case of the honeycomb lattice (figs~\ref{fig:CH}). Calculations of the Chern number for topologically trivial and nontrivial flat bands are shown in figs~\ref{fig:CL}. The phase transition from TI state with $C=-1$ at $E_F=0$ to the trivial metal state is realized via an intermediate topological metal phase. 
 
The calculations of the Chern number in the framework of the Haldane and Lieb models show that the topological metal state is realized as an intermediate state between TI and metal states at doping of system. According to figs~\ref{fig:CH} and figs~\ref{fig:CL}, this topological state is characterized by both ${0<|C|<1}$ and one or two edge mode(s). In~other words, the topological order ${C\neq 0}$ and the chiral edge mode(s) that take place in the metal phase give us possibility to consider this metal state as topological.

\section{Conclusions}

Let us consider the spin variant of the model (\ref{eq:H0}) with different sings of the hopping integrals ${\tau_\sigma =\sigma_z \tau}$~\cite{2}. The system is the sum of two decoupled subsystems of ${\sigma =\uparrow,\downarrow}$ and their Chern numbers $C_\sigma$ are topological indices. Their sum (total Chern number) and the half of their difference (the spin Chern number) define the topological states of such system. A spin current ${J = (\hbar/2e)(J_\uparrow-J_\downarrow)}$ is characterized by a quantized spin Hall conductivity. The edge spin current defines a topological metal state of system, because it is equal to zero in the topologically trivial metal state. The~Chern number in the topological metal state is less than one and nonequal to zero, it defines the topological phase transition from TI state (with ${C=1}$) to the topological metal state (with ${0<C<1}$). The phase transition from the topological metal state to the trivial metal state is realized for ${C\neq 0}$, in this case the jump of the edge charge or spin currents defines the point of topological metal--trivial metal phase transition.

We have show that the topological insulator states in the Lieb lattice are realized at one third and two third band filling when the Fermi level separates narrow (nearly dispersionless) and  neighboring dispersing bands of spinless fermions. The~Chern number of the flat band depends on the parameters of the model, its value is changed at the point of the topological phase transition.

\end{document}